\documentstyle[preprint,aps]{revtex}
\begin{document}
\draft
\title{Coherent stochastic resonance in one dimensional diffusion
with one reflecting and one absorbing boundaries}

\author{Asish K. Dhara and Tapan Mukhopadhyay}
\address{Variable  Energy  Cyclotron  Centre,  1/AF Bidhan Nagar,
Calcutta-700064, India}

\tighten
\date{\today}
\maketitle
\begin{abstract}
It  is  shown  that  the  single-step  periodic  signal (periodic
telegraph signal) can not produce coherent  stochastic  resonance
for  diffusion on a segment with one absorbing and one reflecting
end points while the multi-step periodic signal does. The general
features of this process are exihibited. The  resonant  frequency
is  found to decrease and the mean first passage time at resonant
frequency increases linearly, as we increase the  length  of  the
medium.  The cycle variable is shown to be the proper argument to
express the first passage time density function at  resonance.  A
formula  for  first passage time density function at resonance is
derived in  terms  of  two  universal  functions,  which  clearly
isolates its dependence on the length of the medium.

\end{abstract}
\section{introduction}
After  the  pioneering  achievement  of  separation  of large DNA
molecules in  gel  medium  by  the  application  of  uniform  and
time-dependent   periodic   electric  field  \cite{sch,car},  the
mechanism of cooperative interplay between  random  noise  and  a
deterministic periodic signal attracts considerable interests. It
has  been  found that with this technique, large molecules in the
size range  2  to  400  kb  exihibit  size-dependent  mobilities.
Similar  ideas have also arisen in other types of chromatographic
processes \cite{lin}.

The  first passage time (FPT) is a useful tool to investigate the
diffusive transport property in a medium.  The  theory  of  first
passage  time  has  been  worked  out  in  great  detail for both
infinite  medium  and   explicitly   time-independent   diffusive
processes\cite{wei,gar,van}.      However,     for     explicitly
time-dependent processes and in finite medium an analytic  closed
form  expressions  are  not  available. In this respect also this
problem attracts much attentions to the scientific community.

The  first  analysis of this phenomena has been done for a random
walk on a lattice numerically, and for a diffusive process  in  a
continuous   medium  with  periodic  signal  of  small  amplitude
perturbatively\cite{fle}.  Their  results   indicate   that   the
oscillating field can create a form of coherent motion capable of
reducing  the  first  passage  time by a significant amount. This
enhancement of the mobility of a particle in a  diffusive  medium
by  the  application  of proper oscillating field is known in the
literatures \cite{mas,por} as coherent stochastic resonance(CSR).

In  order to investigate the reason for this cooperative behavior
of random noise and deterministic periodic  signal  this  problem
has  been  formulated  in much simpler terms by approximating the
sinusoidal periodic signal by the telegraph signal\cite{mas}  and
subsequently  it  has been shown\cite{por,dha} that the telegraph
signal can not produce CSR. Finally, when the  sinusiodal  signal
has been approximated as a multi-step periodic signal \cite{dha},
the non-monotonic behavior of mean first passage time (MFPT) with
respect to the characteristic frequency of the periodic signal is
recovered  explaining  the  reason  of  CSR  in  the  case of two
absorbing boundaries explicitly. The general  characteristics  of
the  moments  of  first  passage time density function (FPTDF) in
their calculations \cite{dha}  for  continuous  medium  and  with
arbitrary  amplitide  of  the  periodic signal are found to be in
agreement with the numerical simulation of the random walk  model
on a lattice \cite{fle}.

When  the  phenomenon  of  CSR  is being discussed for the linear
systems, it turns out that the boundary conditions play a crucial
role in some models \cite{doe,bre,git1,git2}. Doering and  Gadoua
have  considered the jumps in a linear double-well potential when
the potential fluctuates between two values at a  rate  $\gamma$.
The  non-monotonic  dependence of the MFPT on $\gamma$ (called ``
the resonance activation") has been found to occur  when  one  of
the   boundaries   is  absorbing  and  the  other  is  reflecting
\cite{doe}. Brey and Cassado-Pascual \cite{bre} have investigated
random walk on a one dimensional lattice in  which  at  any  time
each site has one of two transition rates which are being allowed
to change at random times. Similar effects of boundary conditions
have been observed. Linear models with asymmetric \cite{git1} and
symmetric   \cite{git2}  random  telegraph  signals  (dichotomous
noise) also show that MFPT  behaves  non-monotonically  with  the
jump  rates  for the transition between these two states when one
of the boundaries is absorbing and the other is reflecting.

In this paper we consider an overdamped linear system driven by a
sinusoidal  force  and  embedded  in a noisy environment which is
taken to be Gaussian white.  As  mentioned  before,  no  analytic
closed  form  expression  is available in the literature for this
explicitly time dependent problem. We therefore  approximate  the
sinusoidal   force  (signal)  by  a  multi-step  periodic  signal
(explained below). This is the system which we considered  before
\cite{dha}.  The  only  difference  is that in this paper we have
asymmetric  boundary  conditions,  i.e.,  one  end  point  is   a
reflecting  boundary  and the other one is absorbing while in the
previous work  \cite{dha}  both  the  boundaries  were  taken  as
absorbing.

We  note  that the linear system with single-step periodic signal
(periodic telegraph signal) could be recovered as a special  case
of this system with multi-step periodic signal. The linear system
with single-step periodic signal (periodic telegraph signal) with
one  reflecting  and one absorbing boundaries has been considered
in the literature \cite{git2}.  Non-monotonic  behavior  of  MFPT
with  respect  to  the  characteristic  frequency of the periodic
telegraph signal has been reported \cite{git2}. In this paper  we
show that the telegraph signal does not produce any non-monotonic
behavior  of  MFPT  with  respect to frequency. This result is in
direct contradiction with the result obtained in \cite{git2}. The
explanation of this contradiction is stated in the text below (in
Sec.II and Sec.IIIA).

The  paper  is  organised  as  follows.  In  Sec.II,  we give the
formulation  of  the  problem.  The  basic  structure   of   this
formulation is similar as before \cite{dha}. The only change with
the  previous  one  is  to  incorporate  the effect of asymmetric
boundary conditions. For the sake of completeness we rewrite  the
final   formulae  here  with  incorporation  of  proper  boundary
conditions. After giving the derivation of MFPT  in  Sec.II,  the
results  of  the  calculations are discussed in Sec.III. First we
present the  general  characteristics  of  CSR.  The  calculation
clearly   exhibits   how  resonance  appears  in  our  multi-step
approximation  and  fails  to  show  in   single-step   telegraph
approximation of the periodic signal. The general characteristics
of  the moments and the characteristic features of FPTDF for this
phenomena are also presented in  this  subsection.  In  the  next
subsection  we  focus on the resonance point and demonstrate some
special features associated with it. In particular,  we  show  in
this  subsection (Sec.IIIB) that the cycle variable is the proper
argument to express the FPTDF at resonance. Further, it is  shown
that  the  FPTDF  at  resonance  can be expressed in terms of two
universal functions. This feature clearly isolates its dependence
on the length of the medium. Finally, few concluding remarks have
been added in Sec.IV.

\section{Formulation of the problem}
We  consider  diffusion  in one dimension perturbed by a periodic
force. The motion of  the  particle  is  given  by  the  Langevin
equation

\begin{equation}
\label{eq.1}
\dot{X} = A \sin\Omega t + \xi(t),
\end{equation}
where $X$ refers to the stochastic variable, $A$ and $\Omega$ are
the amplitude and frequency of the sinusoidal signal and $\xi(t)$
is  a  zero  mean  Gaussian  white  noise  of  strength  $D$ with
auto-correlation function given by

\begin{equation}
\label{eq.2}
<\xi(t)\xi(t')> = 2D\delta(t-t').
\end{equation}
The  motion is confined between a reflecting boundary at $x=0$ and
an absorbing  boundary  at  $x=L$.  The  Fokker  Planck  equation
corresponding to Eq.(\ref{eq.1}) is

\begin{equation}
\label{eq.3}
\frac{\partial p(x,t)}{\partial t} = -\frac{\partial j(x,t)}{\partial t} =
- A \sin\Omega t\frac{\partial p(x,t)}
{\partial x} + D\frac{\partial^2 p(x,t)}{\partial x^2},
\end{equation}
where  $p(x,t)$ and $j(x,t)$ refer to the probability density and
probability current density respectively at position $x$  and  at
time $t$. The reflecting boundary condition at $x=0$ implies that
$j(0,t)=0$  and  absorbing  boundary conditions at $x=L$ suggests
that $p(L,t)=0$. We now introduce the dimensionless variables

\begin{equation}
\label{eq.4}
\xi=(A/D)x , \theta = (A^2/D)t , \omega = \Omega/(A^2/D) ,
\end{equation}
to write Eq.(\ref{eq.3}) in terms of new variables:

\begin{equation}
\label{eq.5}
\frac{\partial p(\xi,\theta)}{\partial\theta} = - \sin\omega\theta
\frac{\partial p(\xi,\theta)}{\partial\xi} +\frac{\partial^2p(\xi,\theta)}
{\partial \xi^2}.
\end{equation}
The      boundary      conditions      are      rewritten      as
$j(0,\theta)=p(\Lambda,\theta)=0$, where $\Lambda =  (A/D)L$.  In
the  following  we calculate all the physical quantities in terms
of these new variables and if required, one may translate all the
interpretations  in  terms  of  the  usual   variables   by   the
transformation equations Eq.\ref{eq.4}.

No analytic solution exists for Eq.(\ref{eq.5}) with the boundary
conditions  mentioned.  We thus introduce a scheme to approximate
the force $\sin\omega\theta$  as  a  multi-step  periodic  signal
\cite{dha}.  This  scheme is in contrast to the procedure adopted
in \cite{fle}, where they discretise  the  Eq.(\ref{eq.5})  using
finite difference method and simulate the problem on a lattice of
space and time. The construction of multi-step periodic signal is
as  follows.  We  divide the half cycle of the signal by $(2p+1)$
intervals so that each interval in the horizontal
$\theta$-axis is  of  size  $(\bigtriangleup\theta/(2p+1))$  with
$\omega\bigtriangleup\theta
=\pi$.
We define $(2p+1)$numbers $s_{k}$ along the vertical axis as

\begin{mathletters}
\label{eq.6}
\begin{eqnarray}
s_{k} &=& \frac{[ \sin\frac{k\pi}{2p+1} + \sin\frac{(k-1)\pi}{2p+1} ]}{2}~
; k = 1,2,...,p
\\s_{p+1} &=& 1
\\s_{p+1+r} &=& s_{p+1-r}~ ; r =1,2,...,p .
\end{eqnarray}
\end{mathletters}
Each number $s_k$ is associated with the interval$\frac{(k-1)\bigtriangleup
\theta}{2p+1} <\theta \leq \frac{k\bigtriangleup\theta}{2p+1}$ with
$k = 1,2,...,(2p+1)$. The Eq.(\ref{eq.6}) clearly shows that

\begin{equation}
\label{eq.7}
0<s_{1}<s_{2}<...<s_{p}<s_{p+1}=1>s_{p+2}>s_{p+3}>...>s_{2p+1}>0~.
\end{equation}
Eq.(\ref{eq.7})  states  that in order to reach the maximum value
$(=1)$ of the signal from the zero level we have to have  $(p+1)$
step  up  and  from the maximum to the zero level we have $(p+1)$
step down. This is for the positive half-cycle. For the  negative
half-cycle   similar   constructions  have  been  done  with  the
replacement $s_{k}\rightarrow - s_{k}, \forall k$ and each number
$- s_{k}$ is associated with the  interval  $\bigtriangleup\theta
[1+  \frac{k-1}{2p+1}]<\theta\leq \bigtriangleup \theta [1+ \frac
{k}{2p+1}]$ with $k=1,2,...,(2p+1)$. This  aproximation  for  the
full   one   cycle   of   the  sinusoidal  signal  (as  shown  in
Fig.\ref{fig.1})is then repeated for the next successive  cycles.
The  construction  clearly  shows  that  we  get  back  the usual
telegraph signal with  $p=0$.  Approximation  of  the  sinusoidal
signal  by the usual telegraph signal has been made by \cite{mas}
in the case of two absorbing boundaries.

One  may,  however,  note that the $\omega$ which we have defined
for  this  approximated  signal  is  not  the  same  as  that  of
sinusoidal  signal,  because  the Fourier transform of sinusoidal
signal would give only  one  frequency  while  this  approximated
signal  in  the  Fourier  space  corresponds  to  many sinusoidal
frequencies, especially because of its sharp discontinuities. Yet
we urge this approximation because in  each  small  interval  the
equation becomes time-independent.

The  Fokker-Planck  equation  (Eq.(\ref{eq.5}))  in each interval
with this scheme will be that for a constant bias, namely

\begin{equation}
\label{eq.8}
\frac{\partial p(\xi,\theta)}{\partial\theta} =
\frac{\partial U'(\xi)p(\xi,\theta)}{\partial\xi} +\frac{\partial^2p(\xi,\theta)}
{\partial \xi^2}~,
\end{equation}
where $U'(\xi) = -s (s>0)$ for the positive half cycle and equals
to  $+s  (s>0)$  for  the  negative half cycle with $s$ being the
value of  $s_k$  (Eq.(\ref{eq.6}))  for  the  corresponding  time
interval.  We wish to express the conditional probability density
in  each  interval  in  terms  of  complete  set  of   normalised
eigenfunctions   satisfying  the  boundary  conditions  mentioned
above. For that it is convenient to cast Eq.(\ref{eq.8}) into  an
eigenvalue problem of Schr$\ddot{o}$dinger type by setting

\begin{equation}
\label{eq.9}
p(\xi,\theta) = e^{-\lambda \theta} e^{\pm s\xi /2} \phi(\xi)~.
\end{equation}
Substituting  the  ansatz  Eq.(\ref{eq.9})  in Eq.(\ref{eq.8}) we
obtain

\begin{equation}
\label{eq.10}
\lambda=\mu^2 + s^2/4~,
\end{equation}
\begin{equation}
\label{eq.11}
\phi''+\mu^2 \phi = 0.
\end{equation}
Reflecting boundary condition at $\xi=0$, $j(0, \theta) = 0$, and
absorbing   boundary   condition  at  $\xi=\Lambda$,  $p(\Lambda,
\theta) = 0$, associated with Eq.(\ref{eq.8}) with  the  help  of
Eq.(\ref{eq.9}) take the following form

\begin{equation}
\label{eq.12}
[\frac{1}{2} U' \phi + \phi']_{\xi=0} =0 , \phi(\Lambda)=0.
\end{equation}

In  the  future  development  we  associate the index $n$ for the
positive half-cycle and index $m$ for  the  negative.  Index  $i$
will refer to the cycle number.

Employing  these  boundary  conditions (Eq.(\ref{eq.12})) in some
interval   in   the   positive   half-cycle,    the    normalised
eigenfunctions, $\phi_n(\xi)$ are obtained as

\begin{equation}
\label{eq.13}
\phi_n(\xi) = \left[\frac{2}{\Lambda - \frac{\sin 2\mu_n\Lambda}{2\mu_n}}\right]^{1/2}
\sin\mu_n(\Lambda-\xi)~,
\end{equation}
where $\mu_n$ are obtained by solving the transcendental equation

\begin{equation}
\label{eq.14}
\sin\mu_n\Lambda + \frac{2\mu_n}{s} \cos\mu_n\Lambda =0~,
\end{equation}
with  $s$  being  the  value  of  $s_k$ (Eq.(\ref{eq.6})) for the
corresponding time interval  where  the  conditional  probability
density  is  to  be  decomposed.  The  corresponding  eigenvalues
$\lambda_n$ are given by

\begin{equation}
\label{eq.15}
\lambda_n = \mu_n^2 + s^2/4.
\end{equation}
We  note  that  $\lambda$  in  Eq.(\ref{eq.9})  must be positive.
Therefore in the  range  $-s^2/4<\mu^2<0$,  employing  the  above
procedure    one   obtains   the   eigenfunction   $\phi(\xi)\sim
\sinh\kappa(\xi-\Lambda)$, where $\kappa$ would be obtained as  a
solution of the transcendental equation

\begin{equation}
\label{eq.16}
s\Lambda \sinh\kappa\Lambda + 2\kappa\Lambda\cosh\kappa\Lambda =0~,
\end{equation}
where  $\kappa^2=-\mu^2  ,  \kappa>0.$  As no solution exists for
Eq.(\ref{eq.16}), we say that for  the  positive  half-cycle  the
normalised  eigenfunctions $\phi_n(\xi)$ in Eq.(\ref{eq.13}) form
a complete set. On the other hand, for the  negative  half-cycle,
employing   the   similar   procedure   we  find  the  normalised
eigenfunctions as

\begin{equation}
\label{eq.17}
\phi_m(\xi) = \left[\frac{2}{\Lambda - \frac{\sin 2\mu_m\Lambda}{2\mu_m}}\right]^{1/2}
\sin\mu_m(\Lambda-\xi)~,
\end{equation}
where $\mu_m$ are obtained by solving the transcendental equation

\begin{equation}
\label{eq.18}
\sin\mu_m\Lambda - \frac{2\mu_m}{s} \cos\mu_m\Lambda =0~,
\end{equation}
with  $s$  being  the  value  of $s_k$ for the corresponding time
interval where the  conditional  probability  density  is  to  be
decomposed.  We  note  that Eq.(\ref{eq.17}) and Eq.(\ref{eq.18})
are true for $\mu_m\neq 0.$ For  $\mu_m=0$,  boundary  conditions
suggest  that  one  nontrivial  eigenfunction  exists  only  when
$s\Lambda=2$. Thus  if  the  value  of  $\Lambda$  is  such  that
$s\Lambda=2$,    apart   from   the   eigenfunctions   given   by
Eq.(\ref{eq.17}), one more eigenfunction exists which is

\begin{equation}
\label{eq.19}
\phi_{\mu_m=0}(\xi)=\left(\frac{3}{\Lambda^3}\right)^{1/2}(\xi-\Lambda)~.
\end{equation}
Further,  in  the  range  $-s^2/4<\mu^2<0$,  only  one nontrivial
eigenfunction exists for $s\Lambda>2$. This is given by

\begin{equation}
\label{eq.20}
\phi_{\kappa}(\xi) = \left[\frac{2}{ \frac{\sinh2\kappa\Lambda}{2\kappa}-\Lambda}\right]^{1/2}
\sinh\kappa(\Lambda-\xi)~,
\end{equation}
where  $\kappa^2=-\mu^2,  \kappa>0$ and it is obtained by solving
the transcendental equation

\begin{equation}
\label{eq.21}
-\sinh\kappa\Lambda + \frac{2\kappa}{s}\cosh\kappa\Lambda =0~,
\end{equation}
while  in  the  range $0<s\Lambda<2$, no nontrivial eigenfunction
exists. The corresponding eigenvalues in all the cases are  given
by

\begin{equation}
\label{eq.22}
\lambda_m = \mu_m^2 + s^2/4.
\end{equation}

We  thus  see  that  the  set  of eigenvalues for a given $s$ are
different  in  positive  and  negative  half-cycles  because  the
transcendental  equations, Eq.(\ref{eq.14}) and Eq.(\ref{eq.18}),
are  different.  Thus   the   corresponding   eigenfunctions   of
Eq.(\ref{eq.13})  and  Eq.(\ref{eq.17})  are different functions,
although they have the same form. Apart from  these  differences,
we  have seen that in the range $-s^2/4<\mu^2<0$, there is always
a        nontrivial        eigenvalue        obtained        from
Eqs.(\ref{eq.21})-(\ref{eq.22})  (for  $s\Lambda > 2$) or $s^2/4$
(for $s\Lambda= 2$) and  corresponding  nontrivial  eigenfunction
given   by   Eq.(\ref{eq.20})   (   for   $s\Lambda   >  2$),  or
Eq.(\ref{eq.19}) (for $s\Lambda= 2$ ) in the negative  half-cycle
while in the positive half-cycle, in this particular range, we do
not  have  any  nontrivial  eigenfunction.  For  a given value of
$\Lambda$, the corresponding set of $\{\phi_n(\xi)\}$  forms  the
complete  set  of eigenfunctions in the positive half-cycle while
the   set   $\{\phi_m(\xi)\}$(including    Eq.(\ref{eq.20})    or
Eq.(\ref{eq.19})  as  the  case may be) forms the complete set in
the negative half-cycle.

As  argued  before,  from  our  multi-step  periodic  signal  the
single-step periodic signal (telegraph signal) could be recovered
with $p=0$. In this situation $s$ takes only  one  value,  namely
$s=1$.  The  results  given  above  for  the  eigenfunctions  and
eigenvalues are  true  for  $s=1$  (telegraph  signal)  also.  As
mentioned  in the introduction this special case has been treated
by  Gitterman  \cite{git2}.  However   the   eigenfunctions   and
eigenvalues used there [see Eq.(14) in the paper \cite{git2}] are
wrong  because the eigenvalues do not satisfy the proper boundary
conditions Eq.(\ref{eq.12}) for both the half-cycles.

Once   the   complete   set   of  normalised  eigenfunctions  and
eigenvalues are determined the  conditional  probability  density
function   $p(\xi,\theta  \mid  \xi',\theta')$  in  the
positive half-cycle could be expressed as

\begin{equation}
\label{eq.23}
p(\xi,\theta\mid\xi',\theta') = \sum_{n} u_{n}^+(\xi) u_{n}^-(\xi')
\exp[-\lambda_{n}(\theta-\theta')]~,
\end{equation}
where

\begin{equation}
\label{eq.24}
u_{n}^\pm(\xi) = \exp(\pm s\xi/2) \phi_n(\xi)~ .
\end{equation}
with $s$ as the corresponding value of $s_{k}$ in the appropriate
time   interval   where  the  conditional  probability  is  being
decomposed  and  $\phi_n(\xi)$  and  $\lambda_n$  are  given   by
Eq.(\ref{eq.13})    and    Eq.(\ref{eq.15})   respectively.   The
conditional probability density function  in  any  interval,  say
$l$,  can  then  be  calculated  from  the  previous  history  by
convoluting it in each previous intervals:

\begin{equation}
\label{eq.25}
p(\xi_{l},\theta_{l}\mid \xi_{1},\theta_{1}) =
\int...\int d\xi_{l-1}d\xi_{l-2}...d\xi_{2}\prod_{j=2}^{l}
p(\xi_{j},\theta_{j}\mid \xi_{j-1},\theta_{j-1})~.
\end{equation}
For  the  negative  half-cycle  the  calculation  of  probability
density function is similar except that we have  to  replace  the
index  $n$  by  $m$  and  the  probability  density  function  is
decomposed as

\begin{equation}
\label{eq.26}
p(\xi,\theta\mid\xi',\theta') = \sum_{m} u_{m}^-(\xi) u_{m}^+(\xi')
\exp[-\lambda_{m}(\theta-\theta')]~,
\end{equation}
where  the expressions for $u_{m}^\pm(\xi)$ are
same as in Eq.(\ref{eq.24}) with
$n$ replaced by $m$ and $\lambda_{m}$ are given by Eq.(\ref{eq.22}).

The  survival  probability  at time $\theta$ when the particle is
known to start from $\xi = \xi_0$ at $\theta = 0$ is defined as

\begin{equation}
\label{eq.27}
S(\theta\mid\xi_{0}) =\int_{0}^{\Lambda} d\xi p(\xi,\theta \mid \xi_{0},0)~.
\end{equation}

The  first  passage  time density function (FPTDF) $g(\theta)$ is
defined as

\begin{equation}
\label{eq.28}
g(\theta\mid\xi_{0}) = -\frac{dS(\theta\mid\xi_{0})}{d\theta}~.
\end{equation}
Physically,  $g(\theta)d\theta$  gives  the  probability that the
particle arrives at the absorbing boundary in the  time  interval
$\theta$  and  $\theta + d\theta$. From this density function one
can calculate various moments:

\begin{equation}
\label{eq.29}
<\theta^{j}> = \int_{0}^\infty d\theta \theta ^{j} g(\theta)~.
\end{equation}
From Eq.(\ref{eq.29}) one can easily calculate mean first passage
time(MFPT)    $<\theta>$    and    the   variance   $\sigma^2   =
<\theta^2>-<\theta>^2$ of the density function $g(\theta)$.

It  is  then  quite  straight-forward  to  calculate the survival
probability at any interval of any cycle. We will write down  the
final formulae:

\begin{mathletters}
\label{eq.30}
\begin{eqnarray}
S_{i, k}^+(\theta\mid\xi_{0})=&& C^+_{n_{(2p+1)(i-1)+k}} \nonumber \\
&& \times \exp[-\lambda_{n_{(2p+1)(i-1)+k}}(\theta-2(i-1)\bigtriangleup\theta)] \nonumber \\
&& \times O_{k}(u^-_{n_{(2p+1)(i-1)+k}})~,\nonumber\\
&&;[2(i-1)+\frac{k-1}{2p+1}]\bigtriangleup\theta
<\theta\leq[2(i-1)+\frac{k}{2p+1}]\bigtriangleup\theta~,\nonumber\\
&&;k=1,2,...,(2p+1)~,
\end{eqnarray}

\begin{eqnarray}
S_{i, k}^-(\theta\mid\xi_{0})=&& C^-_{m_{(2p+1)(i-1)+k}} \nonumber \\
&& \times \exp[-\lambda_{m_{(2p+1)(i-1)+k}}(\theta-(2i-1)\bigtriangleup\theta)] \nonumber \\
&& \times E_{k}(u^+_{m_{(2p+1)(i-1)+k}})~, \nonumber\\
&&;[(2i-1)+\frac{k-1}{2p+1}]\bigtriangleup\theta
<\theta\leq[(2i-1)+\frac{k}{2p+1}]\bigtriangleup\theta~, \nonumber\\
&&;k=1,2,...,(2p+1)~,
\end{eqnarray}
\end{mathletters}
where

\begin{mathletters}
\label{eq.31}
\begin{eqnarray}
C_{n}^+ =&&\int_{0}^{\Lambda} d\xi u_{n}^+(\xi)~,
\end{eqnarray}

\begin{eqnarray}
C_{m}^- =&&\int_{0}^{\Lambda} d\xi u_{m}^-(\xi)~,
\end{eqnarray}
\end{mathletters}
and  the functions $O_k, E_k$ are generated through the recursion
relations:

\begin{mathletters}
\label{eq.32}
\begin{eqnarray}
O_1(u^-_{n_{(2p+1)(i-1)+1}})=F_{i-1}(u^-_{n_{(2p+1)(i-1)+1}})~,
\end{eqnarray}

\begin{eqnarray}
O_k(u^-_{n_{(2p+1)(i-1)+k}})=
&&<u^-_{n_{(2p+1)(i-1)+k}}\mid u^+_{n_{(2p+1)(i-1)+(k-1)}}>\nonumber\\
&&\times \exp[\frac{(k-1)\bigtriangleup\theta}{2p+1}(\lambda_{n_{(2p+1)(i-1)+k}}
-\lambda_{n_{(2p+1)(i-1)+(k-1)}})]\nonumber\\
&&\times O_{k-1}(u^-_{n_{(2p+1)(i-1)+(k-1)}})~, \nonumber\\
&&;k=2,3,...,(2p+1)~,
\end{eqnarray}

\begin{eqnarray}
E_1(u^+_{m_{(2p+1)(i-1)+1}})=
&&<u^+_{m_{(2p+1)(i-1)+1}}\mid u^+_{n_{(2p+1)i}}>\nonumber\\
&&\times \exp[-\bigtriangleup\theta \lambda_{n_{(2p+1)i}}]\nonumber\\
&&\times O_{2p+1}(u^-_{n_{(2p+1)i}})~,
\end{eqnarray}

\begin{eqnarray}
E_k(u^+_{m_{(2p+1)(i-1)+k}})=
&&<u^+_{m_{(2p+1)(i-1)+k}}\mid u^-_{m_{(2p+1)(i-1)+(k-1)}}>\nonumber\\
&&\times \exp[\frac{(k-1)\bigtriangleup\theta}{2p+1}(\lambda_{m_{(2p+1)(i-1)+k}}
-\lambda_{m_{(2p+1)(i-1)+(k-1)}})]\nonumber\\
&&\times E_{k-1}(u^+_{m_{(2p+1)(i-1)+(k-1)}})~,\nonumber\\
&&;k=2,3,...,(2p+1)~,
\end{eqnarray}

\begin{eqnarray}
F_i(u^-_{n_{(2p+1)i+1}})=
&&<u^-_{n_{(2p+1)i+1}}\mid u^-_{m_{(2p+1)i}}>\nonumber\\
&&\times \exp[-\bigtriangleup\theta \lambda_{m_{(2p+1)i}}]\nonumber\\
&&\times E_{2p+1}(u^+_{m_{(2p+1)i}})~,
\end{eqnarray}
\end{mathletters}
with   $F_{0}(u^-_{n_{1}})  =  u^-_{n_{1}}(\xi_0)$.  The  angular
bracket in any equation implies dot product of the  corresponding
functions, for e.g.,

\begin{equation}
\label{eq.33}
<u^+\mid u^-> = \int_{0}^{\Lambda} d\xi u^+(\xi)u^-(\xi)~.
\end{equation}
The  cycle  variable  $i$  runs  over  positive  integers;  i.e.,
$i=1,2,3,...$. The positive and negative symbols of the  survival
probabilities  indicate its value over positive and negative part
of the cycles  respectively.  In  all  these  expressions,  viz.,
Eqs.(\ref{eq.30})-(\ref{eq.32}),  any subscript either $n$ or $m$
or both whereever they appear more than once the  summation  over
them  are  implied.  The  effect  of  history  is explicit in the
expressions  for  survival  probabilities.  Once   the   survival
probability $S(\theta\mid\xi_0)$ is obtained from these formluae,
the  FPTDF,  MFPT  and the corresponding variance are obtained by
employing Eqs.(\ref{eq.28})-(\ref{eq.29}). Evaluation of MFPT and
other relevant quantities requires sum of infinite  series  which
must  be truncated in order to obtain a final result. Convergence
of MFPT is ensured by gradually increasing the  number  of  terms
(i.e., number of eigenvalues) for the calculation. The process is
truncated  when  MFPT  does  not change upto two decimal point of
accuracy with the change of number of terms.

\section{results and discussions}
The   survival   probability,   mean  first  passage  time(MFPT),
corresponding  variances   and   first   passage   time   density
functions(FPTDF)  are  calculated  using the derived formulae for
this process. The results are summarised below.

\subsection{General features of CSR}
The  MFPT  is calculated for single-step telegraph signal $(p=0)$
with $\xi_0=\Lambda/2$. Most of the calculations  are  done  with
this specific value of $\xi_0$. The variation of the results with
variation  of  $\xi_0$ is also demonstrated [see the text below].
No nonmonotonous behavior is observed in  MFPT  as  we  vary  the
frequency   $\omega$.  This  is  in  complete  disagreement  with
Gitterman's observation\cite{git2}. Because of the wrong  set  of
eigenvalues and corresponding eigenfunctions in both positive and
negative  half-cycles used to express the conditional probability
density, Gitterman \cite{git2} observed  nonmonotonic  dependence
of  MFPT  with  respect  to  the  characteristic frequency of the
periodic telegraph signal.

The  calculation  is  done  for the length $\Lambda = 20$ and the
result is shown in the curve $(a)$ of  Fig.2.  However,  when  we
take  $p=1$,  i.e.,  when the sinusodal signal is approximated by
two-step periodic signal, the calculation of MFPT  for  the  same
length shows clearly the nonmonotonous behavior. This is shown in
curve  $(b)$ of the same figure. This result clearly demonstrates
that mere flipping of the bias  (signal)  direction  periodically
would  not  produce  the coherent motion. As the rate of flipping
increases it merely prevents  the  particle  more  to  reach  the
absorbing boundary and therefore MFPT increases monotonically. It
may be noted that when the flipping rate is very high, the effect
of  signal  is  almost  nil  and  the  transport  is  effectively
diffusive in nature. This is  of  course  true  in  any  type  of
periodic  signal. Therefore, for any type of approximation of the
sinusoidal signal or for any value of  $p$,  this  feature  would
show up. In particular, for $p=1$, we observe from curve $(b)$ of
fig.2  that  MFPT  would asymptotically reach the diffusive limit
$3\Lambda^2/8$ (= 150 in this case). The usual  telegraph  signal
offers  a  constant bias of maximum magnitude for the larger time
than for a two-step approximation. Hence the particle always  has
a  larger probability of reaching the absorbing boundary in short
time for $p=0$ case than for $p>0$ case. Hence MFPT for $p=0$ and
for any $\omega$ is always less than  for  $p>0$  case.  This  is
observed in Fig.2.

The  application of any bias always reduces the MFPT than for the
non-biased diffusion. In CSR we always have a competition between
diffusion and oscillatory effect of  the  bias.  For  very  large
frequency  as  the  bias  effect  becomes  ineffective MFPT would
essentially be guided by diffusive process. For zero frequency of
the multi-step periodic  signal  the  MFPT  can  be  analytically
evaluated.  When  it  starts  from the mid-point of the medium it
expresses as

\begin{equation}
\label{eq.34}
 <\theta(\omega=0,\xi_0=\Lambda/2)>
=0.5(\Lambda/s_1)-\left(\frac{2e^{-3s_1\Lambda/4}}{s_1^2}\right)\sinh(s_1\Lambda/4)~.
\end{equation}
When  frequency  is  very  small,  the  process  is predominantly
diffusion with constant  value  $s_1=0.5  \sin(\frac{\pi}{2p+1})$
effective  for $0<\theta\le\frac{\pi}{\omega(2p+1)}$. However, as
frequency increases slowly, the probability of  having  increased
bias  value  $s_2$ (=1 for $p=1$) before it reaches the absorbing
boundary  increases.  This  bias  force  reduces   the   survival
probability  and  also  MFPT. Hence one would expect a minimum to
MFPT. On the otherhand, for usual telegraph signal ($p=0$  case),
for  very  low  frequency,  from  the  very  beginning bias force
affects  the  particle  with  its  maximum  strength.  When   the
frequency is very low, this constant bias diffusion continues for
a longer time and there is no change-over of the magnitude of the
bias  as  in the case of $p=1$. After having a flip, the particle
again suffers a constant bias diffusion in the direction opposite
to the previous one  and  towards  the  reflecting  boundary.  As
frequency  increases slowly, this picture remains unchanged until
a stage reaches for which the flipping  effect  becomes  dominant
during  the  particle's  survivality  inside  the medium and MFPT
increases. This is observed in Fig.2.

Next we continue all our calculation with $p=5$ or, with six-step
telegraph signal. Calculation reveals that the value of MFPT does
not  change  much  from that with $p=1$. On the other hand, $p=5$
signal  approximates  the  sinusidal  signal  better  than  $p=1$
signal.  We  restrict our calculation with $p=5$ approximation of
the periodic signal.

We calculate the MFPT $<\theta>$ and the variance $\sigma^2$ as a
function  of frequency $\omega$ for different lengths $(\Lambda =
10,20,30,40,50)$.  These  are  presented  in  Fig.3   and   Fig.4
respectively.   Both  the  cumulants  go  through  a  minimum  as
frequency rises from very low value for each length $\Lambda$. It
is observed that the minimum for both the moments occur almost at
the same frequency for each length. This shows that  the  maximum
cooperation  between  the  deterministic  signal and random noise
occurs at this resonant frequency. The value of  MFPT  $<\theta>$
increases   with   the   length   at  all  frequencies.  This  is
understandable because as length increases,  on  an  average  the
particle  will  spend more time in the medium before reaching the
absorbing boundary. It is also observed  that  the  frequency  at
which  the  minimum  occurs  shift  towards  low frequency as the
length increases.

Fig.4  demonstrates  the  lowering  of the dispersion at resonant
frequencies confirming that the cooperation is maximum  at  these
frequencies.  Dispersion  is  more for higher lengths and as seen
from the figure the dispersion merges to a specific value at very
low frequency at various lengths.

All  the  previous calculations are done when the particle starts
initially from the mid point of the medium,  i.e.,  $\xi_{0}$  in
Eq.(\ref{eq.30})   is   taken   as  $\Lambda/2$.  At  the  length
$\Lambda=30$ the resonant frequency is  found  to  be  0.06.  The
calculations  are done one at resonant frequency and other two at
the off- resonant  frequencies  ($\omega=0.1$  and  $\omega=0.0$)
when   the  particle  starts  from  $\xi_{0}=\beta\Lambda$  where
$\beta$ lies between 0 and 1. For zero frequency the MFPT can  be
analytically obtained. Its expression reads as

\begin{equation}
\label{eq.35}
 <\theta(\omega=0, \beta\Lambda)>
=\frac{\Lambda (1-\beta)}{s_1}-\left(\frac{2e^{-s_1\Lambda (1+\beta)/2}}{s_1^2}\right)
\sinh(s_1\Lambda (1-\beta)/2)~.
\end{equation}
The  curves  are  shown in Fig.5. It is evident that the value of
$<\theta>$ is less for resonant frequency (curve $(a)$) than  for
its  value  for off-resonant frequencies(curves $(b)$ and $(c)$).
For each curve the maximum value of $<\theta>$  occurs  at  lower
values  of  $\beta$ or, when the particle starts from the left of
the interval (near the reflecting boundary).  Our  signal  starts
with  positive  half-cycle and therefore the survival time of the
particle would be more if the particle starts from  the  left  of
the  interval. On the other hand, if it starts from right half of
the medium (near the absorbing boundary), the  initial  surge  of
the  signal  helps  the  particle to reach the absorbing boundary
more quickly. Hence average time of duration decreases. For curve
(b) where the frequency of the signal $(\omega =  0.1)$  is  more
than the resonant frequency, the oscillatory contribution is more
than  that  for the resonant frequency making the MFPT large than
for curve $(a)$. For curve $(c)$  where  the  frequency  is  zero
$(\omega  =  0.0)$,  the  particle  experiences  a  constant bias
$(s=s_1)$ towards the absorbing boundary all the time. This is in
contrast to  the  other  two  cases  where  the  particle  has  a
probability  to  experience bias of magnitude more than $s_1$. As
the value of $s_1$ (the first value of six-step periodic  signal)
is  smallest  hence  the curve $(c)$ lies always above the curves
$(a)$ or $(b)$.

The  FPTDF  $g(\theta;  \omega)$  for  different  frequencies are
calculated and plotted as a function  of  $\theta$  for  a  given
length  $\Lambda=30$  [Figs.6a  \&  6b].  One can clearly see the
evolution of the FPTDF profile as the  frequency  increases  from
$\omega=0.001$ to $\omega=0.1$.

At  very  small  frequency (e.g. $\omega=0.001$), the particle is
acted on by a constant force $s=s_1$ (in the  multistep  periodic
approximation)   almost  all  the  time  before  it  reaches  the
absorbing boundary. The FPTDF curve shows that we have  only  one
maximum in the entire $\theta$-range.

As  the  frequency  slowly  increases, one finds that, apart from
only one maximum (like the one at  very  small  frequency)  other
small  peaks  at  larger  $\theta$  also show up gradually. Also,
these new small peaks start becoming stronger  as  the  frequency
increases.  This  is  because the particle which initially sees a
constant bias ($s=s_1$) for some  time  encounters  an  increased
bias  after  a  while (larger $\theta$) and so on before reaching
the absorbing boundary. As $\int{g(\theta)  d\theta}  =  1$,  the
area  under  the  major  profile  decreases and is compensated by
extra peaks at higher $\theta$. We  therefore  see  that  as  the
frequency increases continuously, the small peaks that show up at
higher $\theta$ start growing and at the same time the first peak
slowly decreases keeping the total area same.

The  resonance occurs at $\omega=0.06$ for $\Lambda=30$, as shown
in Fig.6b. The MFPT has a minimum at that frequency. It has  been
argued  that the synchronization between the signal and the noise
is maximum at this  frequency  causing  the  enhancement  of  the
probability  of  reaching the absorbing boundary at a short time.
The FPTDF profile at  resonance  (at  this  frequency)  evidently
shows  a  dominant  single peak. We also see that the constituent
profiles of FPTDF adjust themselves in a very  distinct  way,  as
the  frequency  increases, in order to produce a dominant maximum
at the resonance.

Beyond  the resonance (at higher frequencies) the peaks are quite
numerous, distinct and identified separately, while the  area  of
the  dominant  peak  (at  resonance)  starts decreasing. In other
words, the degree of synchronization is getting reduced as we  go
beyond
the resonant frequency -- if we identify the area of the peaks
as  the  degree  of  synchronization  (considering  total area is
normalized).

\subsection{Special features at resonance}
In  this subsection we concentrate on the behaviour of the system
at the resonance point. We have already  discussed  some  general
characteristics  of  CSR in the previous subsection. We find that
for each length, $\Lambda$, a corresponding frequency  $\omega^*$
exists   for  which  $<\theta>$  and  $\sigma^2$  become  minimum
implying that the maximum cooperation between  the  deterministic
periodic  signal  and  random  noise of the environment is taking
place in helping the particle to reach  the  absorbing  boundary.
One  therefore  would  naturally  inquire  about  the relation of
$\omega^*$ with $\Lambda$. The curve of $\omega^*$ as a  function
of  $\Lambda$  is  plotted in Fig.7. In the range of $\Lambda$ we
studied  this  curve  is  very  well  fitted  with  the   formula

\begin{equation}
\label{eq.36}
\omega^*  = C/\Lambda^{\gamma}~,
\end{equation}
where $C=0.8053$, and $\gamma=0.7615$.

The  values  of MFPT at resonance $<\theta(\omega^*)>$ is plotted
against the length $\Lambda$ in Fig.8 and  within  the  range  of
$\Lambda$ we consider the relation between them is fitted to

\begin{equation}
\label{eq.37}
<\theta(\omega^*)> = 0.79 \Lambda + 1.85~.
\end{equation}
Of  course, there will be deviation from this linear behaviour as
$\Lambda$ decreases further because  $<\theta>$  can  not  become
positive  for  $\Lambda=0$(corresponding  to  $L=0$);  $<\theta>$
should be zero at $\Lambda=0$.

We  have  already  seen  that  at the resonance frequency we have
single dominant peak of FPTDF, $g(\theta)$[Fig.6b]. Since it is a
general feature, for each length $\Lambda$  we  should  get  such
behaviour.  When  we  plot  $g(\theta)/\omega^*$ as a function of
$[\omega^*(\Lambda)\theta]$,  we  find  that   the   curves   for
different  lengths  superpose  over  each  other  [Fig.9] and the
pattern  of  $g(\theta)/\omega^*$  for  different  $\Lambda$   or
$\omega^*$  is  very  similar,  i.e.,  at  particular  values  of
$[\omega^*\theta]$, all curves show their maxima, and  change  in
the  behavioral  patterns of the curves occur exactly at the same
places of$[\omega^*\theta]$. This shows  that  $[\omega^*\theta]$
or  the  cycle  number  is  the  correct variable to describe the
resonance behaviour. We may further note  that  such  scaling  of
FPTDF  would  not  be  possible  for any frequency other than the
resonant frequencies because  any  frequency  which  is  not  the
resonant frequency for one length may turn out to be the resonant
frequency  for  some  other  length and the features of FPTDF are
different for resonant and off-resonant frequencies as  has  been
observed  from Fig.6a - Fig.6b. The major dominant peaks of FPTDF
$g(\theta)/\omega^*$ for different lengths$(\Lambda = 20, 25, 30,
35, 40, 50)$ are drawn as a  function  of  $[\omega^*\theta]$  in
Fig.9.  The  lowermost  curve is for $\Lambda = 20$ and as length
increases the upper curves are generated. The peaks for  all  the
curves occur nearly at a quarter of a cycle.

Having  found  the proper scaling of the argument of FPTDF, it is
natural   to   enquire   whether    the    FPTDF,    $f    \equiv
g(\theta)/\omega^*$  can  also  be  scaled properly, so that once
$f(x)$ with $x=\omega^* \theta$ is  found  for  one  length,  the
function $f(x)$ can be obtained for any arbitrary length.

In  order  to investigate this issue, first we observe that going
from $f(x;\Lambda_1)$ to $f(x;\Lambda_2)$, where $\Lambda_1$  and
$\Lambda_2$  are  two  different  lengths we have to multiply one
function by different  amount  depending  on  the  value  of  the
argument. Thus if at all any scaling relation exists, this should
be of the form

\begin{equation}
\label{eq.38}
f_{\mu\Lambda}(x) = f_{\Lambda}(x) \mu^{\alpha(x)}
\end{equation}

It  is  indeed found to be true with $\alpha(x)$ given in Fig.10.
We note that $\alpha(x)$ is universal. From the relation (38)  it
is obvious that

\begin{equation}
\label{eq.39}
\Lambda^{-\alpha(x)} f_{\Lambda}(x) = [\mu\Lambda]^{-\alpha(x)} f_{\mu\Lambda}(x).
\end{equation}
The      Eq.(\ref{eq.39})     shows     that     the     function
$\Lambda^{-\alpha(x)} f_{\Lambda}(x)$ is independent of $\Lambda$
and  it  is  again  universal.  We  verify  this  result  in  our
calculation of FPTDF for different lengths and call this function
$U(x)$.  The  $lnU(x)$  is  also plotted in Fig.10. Thus we could
express $f_{\Lambda}(x)$ at resonance in terms of  two  universal
functions $U(x)$ and $\alpha(x)$ as

\begin{equation}
\label{eq.40}
f_{\Lambda}(x)= U(x) \Lambda^{\alpha(x)}
\end{equation}
The  expression(40)  clearly  isolates the dependence of FPTDF on
length  of  the  medium  $\Lambda$   at   resonance.   The   form
(\ref{eq.40})  for  $f_{\Lambda}(x)$  is  plotted  for  a typical
length $\Lambda=30$ and  compared  with  actual  data  points  in
Fig.11.  It  is  giving an excellent agreement. Further with this
form, the MFPT is found for different $\Lambda$. It  also  yields
required  behaviour  as  in Eq.(\ref{eq.37}). This shows that not
only we have found a scaling behaviour given by  Eq.(\ref{eq.38})
which  takes  $f(x)$  from  one  length  to the other, it is also
possible to  obtain  a  simple  expression  Eq.(\ref{eq.40})  for
$f_{\Lambda}(x)$ for any arbitrary length.

\section{concluding remarks}
We consider a diffusive transport process perturbed by a periodic
signal in continuous one dimensional medium having one reflecting
and  one  absorbing  boundaries.  We  showed  explicitly that the
cooperative behaviour between the deterministic  periodic  signal
and  random  noise  leading  to  coherent  motion occurs when the
time-dependent sinusoidal signal is approximated by  a  multistep
periodic signal and not with single-step telegraph signal.

Although  we study the process with six-step periodic signal, the
formulation is quite general and applicable for any approximation
with arbitrary number of steps.  This  formulation  can  also  be
applied  to any arbitrary continuous periodic signal. Further, no
perturbation  of  the  signal  amplitude  is  assumed   in   this
formulation.

It  is  observed  that  for  large time oscillation of the signal
plays a dominant role in the transport  while  in  the  low  time
regime  frequency dependent bias force (i.e.,the chance of having
increased values of the bias in the same direction  is  more  for
high  frequency  than  for low frequency) has the key factor. For
very high frequency the bias effect is practically absent and the
motion is purely  diffusive  in  nature.  At  the  resonance  the
maximum  cooperation  between  the  noise and the periodic signal
takes place making the MFPT a minimum.

The  resonant frequency decreases and the mean first passage time
at resonant frequency increases  linearly,  as  we  increase  the
length  of  the medium. The corresponding relations are obtained.
This fact provides a measure of the time scale of this process at
resonance.

Other  important  characteristic  that  we observe is that at the
resonance the FPTDF for various lengths have similar behaviour as
a function of cycle number. The curve shows that there is  single
dominant  peak,  which  is  a  reflection  of  the  fact that the
synchronization between the deterministic periodic signal and the
random noise is maximum at resonance. The peak positions of these
curves occur very near to a quarter of a cycle showing  that  the
cycle  number  is  the  correct  argument  to  describe  FPTDF at
resonance.

We  also  show that there exists a scaling relation between FPTDF
at various lengths through some universal  function  $\alpha(x)$.
The  exact expression for FPTDF at resonance is obtained in terms
of two universal functions, which clearly isolates its dependence
on the length of  the  medium.  This  is  a  special  feature  of
coherent  stochastic  resonance.  This  form  may  be  of  use in
obtaining quicker result in cases where more complex situation is
called for.

There  is  also slight discrepancy in the position of the minimum
of $\sigma^2$ in comparison to the minima of  $<\theta>$  [Fig.3,
Fig.4].  This  may  be  due to the fact that all calculations are
made to an end  when  the  survival  probability  takes  a  value
$1\times 10^{-3}$. We observe that if we cut off the calculations
for  more lower values of survival probability it does not affect
MFPT but  the  variances  are  slightly  affected.  Also  if  one
approximates  the  sinusoidal  signal  better,  by  more  than  a
six-step periodic signal one could obtain the  positions  of  the
minima  of  variances  at  exactly  the same places as those with
MFPT.

From  Fig.9  we  observe a slight deviation of the peak positions
but we believe that if the sinusoidal signal is  approximated  by
more  than  a  six-step  periodic  signal the position of all the
peaks will be the same.

\begin{figure}
\caption{Sinusoidal   signal(dashed   curve)   and   approximated
six-step $(p=5)$ periodic signal(solid curve)  for  the  full  one
cycle as a function of $\theta$.}
\label{fig.1}
\end{figure}

\begin{figure}
\caption{MFPT $<\theta>$ as a function of $\omega$ ;
(a)for p=0 ;the usual telegraph
signal (b) for p=1 ; the two-step periodic signal,
[$\Lambda=20,\xi_0=\Lambda/2$].}
\label{fig.2}
\end{figure}

\begin{figure}
\caption{MFPT  $<\theta(\omega)>$  as  a  function  of  frequency
$\omega$  ;  (a)$\Lambda=10$,  (b)$\Lambda=20$,  (c)$\Lambda=30$,
(d)$\Lambda=40$, (e)$\Lambda=50$, $[p=5,\xi_0=\Lambda/2]$.}
\label{fig.3}
\end{figure}

\begin{figure}
\caption{The  variance  $\sigma^2$  as  a  function of $\omega$ ;
(a)$\Lambda=10$,        (b)$\Lambda=20$,         (c)$\Lambda=30$,
(d)$\Lambda=40$, (e)$\Lambda=50$, $[p=5,\xi_0=\Lambda/2]$.}
\label{fig.4}
\end{figure}

\begin{figure}
\caption{MFPT $<\theta>$ as a function of $\beta$ for length $\Lambda=30$ ;
(a) for resonant frequncy $\omega^* = 0.06$
(b) for off-resonant frequncy $\omega = 0.1$
(c) for off-resonant frequncy $\omega = 0.0$,
$[p=5]$. }
\label{fig.5}
\end{figure}

\begin{figure}
\caption{(a) FPTDF $g(\theta)$ as a function of $\theta$ for $\Lambda=30$
before resonance for frequencies $\omega$ = 0.001, 0.003, 0.005, 0.01
(b) FPTDF $g(\theta)$ as a function of $\theta$ for $\Lambda=30$
on and after resonance for frequencies $\omega = 0.06$(resonant),
$\omega =0.07, 0.09, 0.1 $ respectively, $[p=5,\xi_0=\Lambda/2]$.}
\label{fig.6}
\end{figure}

\begin{figure}
\caption{Resonant  frequency  $\omega^*$  as a function of length
$\Lambda$.}
\label{fig.7}
\end{figure}

\begin{figure}
\caption{MFPT at the resonant frequency $<\theta(\omega^*)>$ as a
function of length $\Lambda$. }
\label{fig.8}
\end{figure}

\begin{figure}
\caption{The  dominant  peaks of $g(\theta)/\omega^*$ at resonant
frequencies for different lengths ($\Lambda = 20, 25, 30, 35, 40,
50$)  as  a  function  of $\omega^*\theta$. The lowermost
curve is for $\Lambda = 20$ , and as length  increases  gradually
upper curves are generated, $[p=5,\xi_0=\Lambda/2]$.}
\label{fig.9}
\end{figure}

\begin{figure}
\caption{The   universal   functions  $\alpha(x),  lnU(x)$  as  a
function of $x$. }
\label{fig.10}
\end{figure}

\begin{figure}
\caption{$f_{\Lambda}(x)$   as   a   function   of   $x$   for
$\Lambda=30$ at resonance from actual data points (filled circle)
$[p=5,\xi_0=\Lambda/2]$
and $f_{\Lambda}(x)$  according  to the formula (40) as a function
of $x$ for $\Lambda=30$(solid curve).}
\label{fig.11}
\end{figure}


\begin{references}

\bibitem{sch}  D.  C. Schwartz and C. R. Cantor, Cell {\bf37}, 67
(1984).

\bibitem{car}  G.  F.  Carle,  M. Frank, and M. V. Olson, Science
{\bf232}, 65 (1986).

\bibitem{lin} I. J. Lin and L. Benguigi, Sep. Sci. Tech. {\bf20},
359 (1985).

\bibitem{wei} G. H. Weiss, Adv. Chem. Phys. {\bf13}, 1 (1967).

\bibitem{gar}  C.  W.  Gardiner,  {\it  A  handbook of Stochastic
Methods}, 2nd ed. (Springer-Verlag, New York, 1985).

\bibitem{van}  N.  G.  van  Kampen,  {\it Stochastic processes in
physics and chemistry} (North-Holland, Amsterdam, 1991).

\bibitem{fle}  J.  E.  Fletcher,  S.  Havlin, and G. H. Weiss, J.
Stat. Phys. {\bf51}, 215 (1988).

\bibitem{mas}  J.  Masoliver, A. Robinson, and G. H. Weiss, Phys.
Rev. E{\bf51}, 4021 (1995).

\bibitem{por} J. M. Porra, Phys. Rev. E{\bf55}, 6533 (1997).

\bibitem{dha}  Asish  K. Dhara and Tapan Mukhopadhyay, Phys. Rev.
E{\bf60}, 2727 (1999).

\bibitem{doe}  C.  R.  Doering and J. C. Gadoua, Phys. Rev. Lett.
{\bf69}, 2318 (1992).

\bibitem{bre}  J.  J.  Brey  and  J.  Cassado-Pascual,  Physica  A
{\bf212}, 123 (1994).

\bibitem  {git1}  M.  Gitterman,  R. I. Shrager, and G. H. Weiss,
Phys. Rev. E{\bf56}, 3713 (1997).

\bibitem {git2} M. Gitterman, Phys. Rev. E {\bf61}, 4726 (2000).

\end{references}
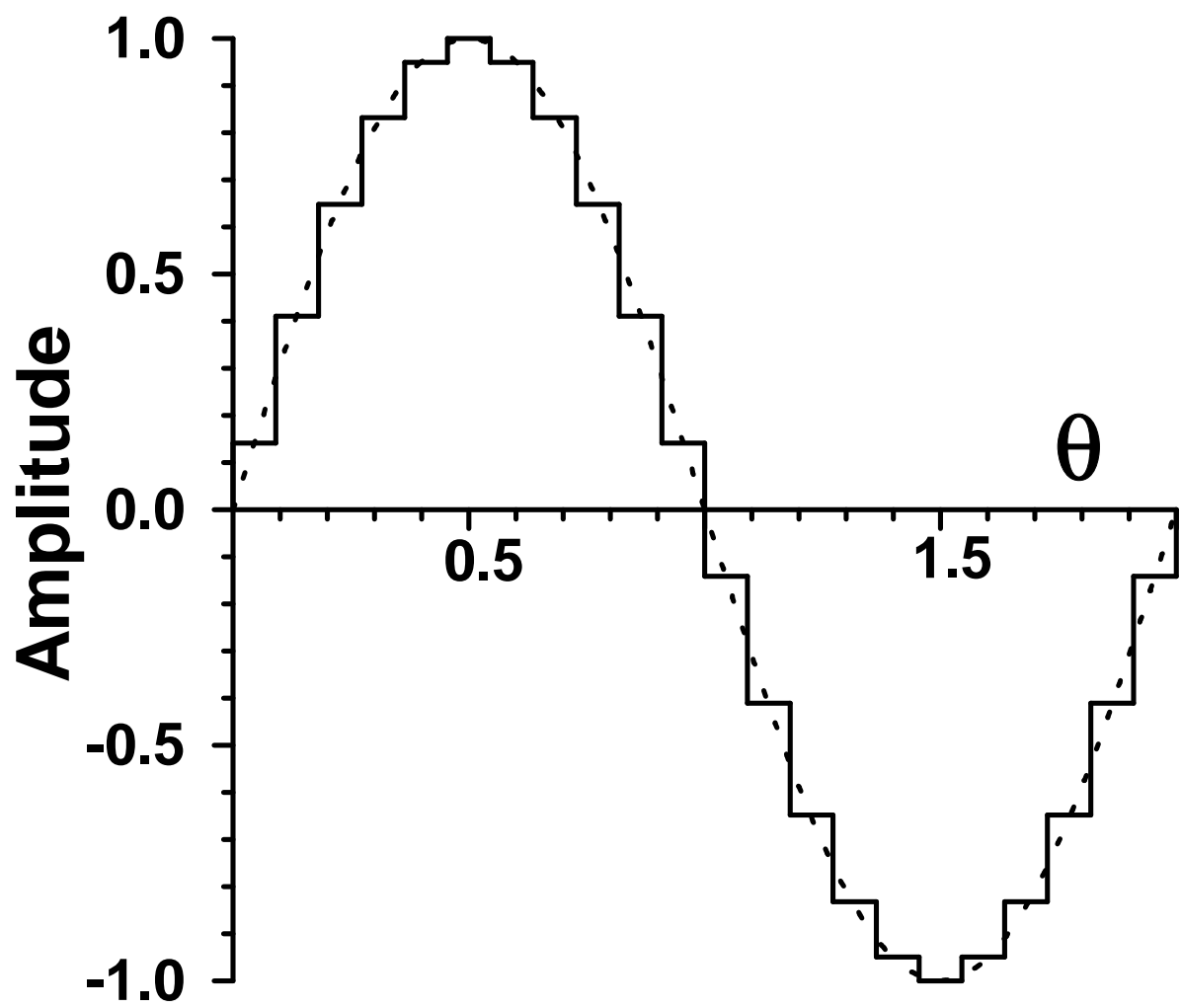
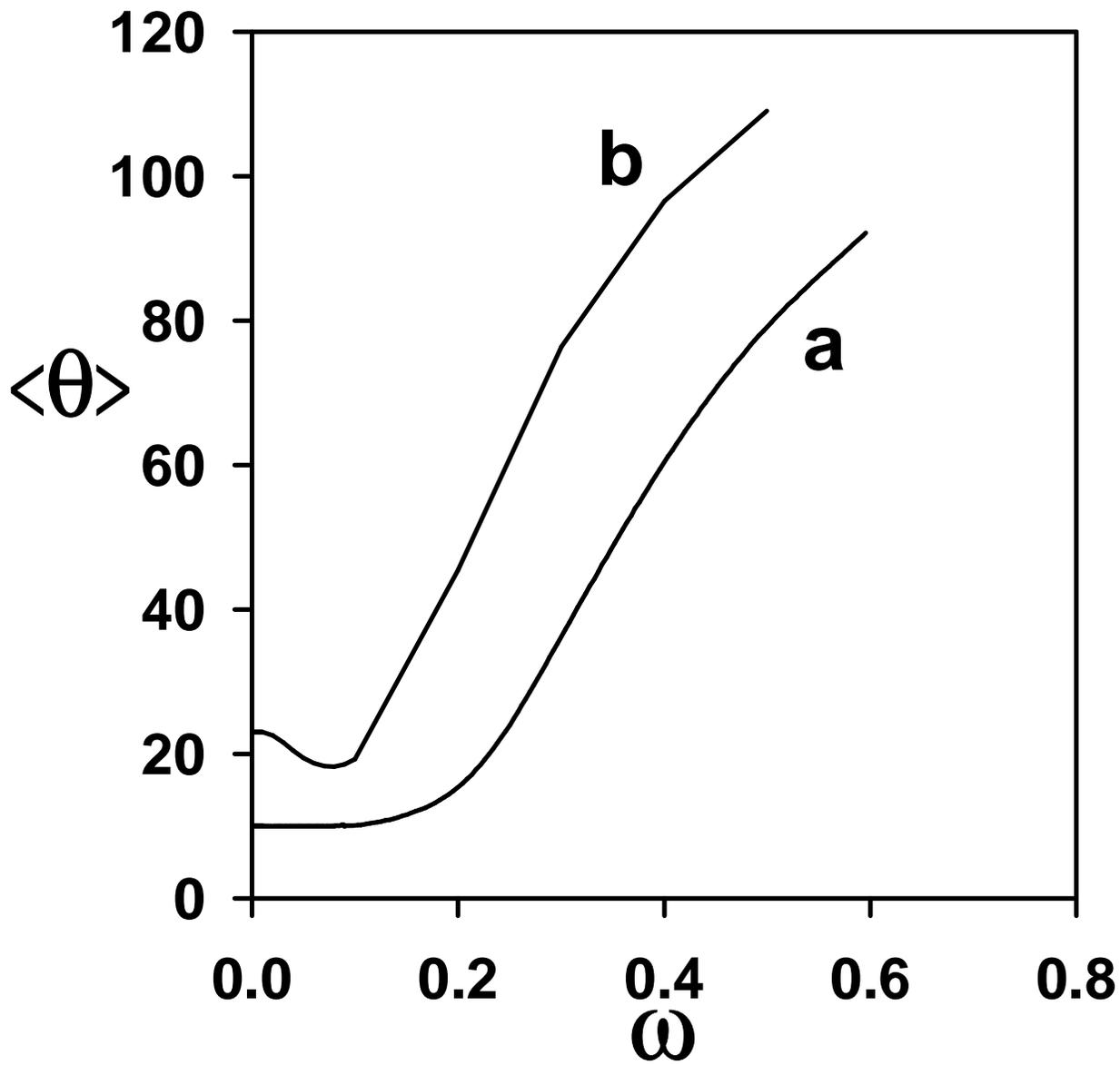
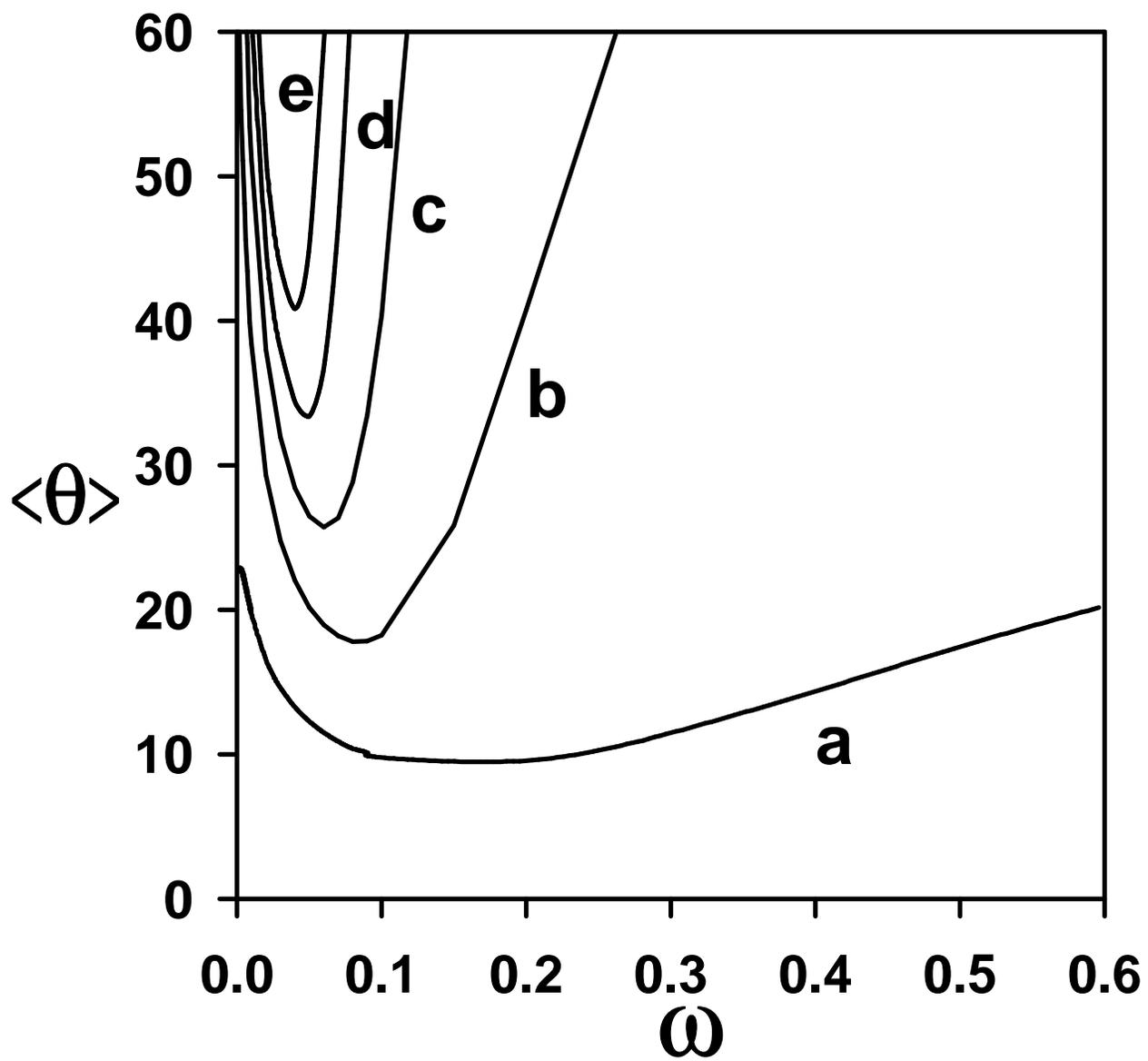
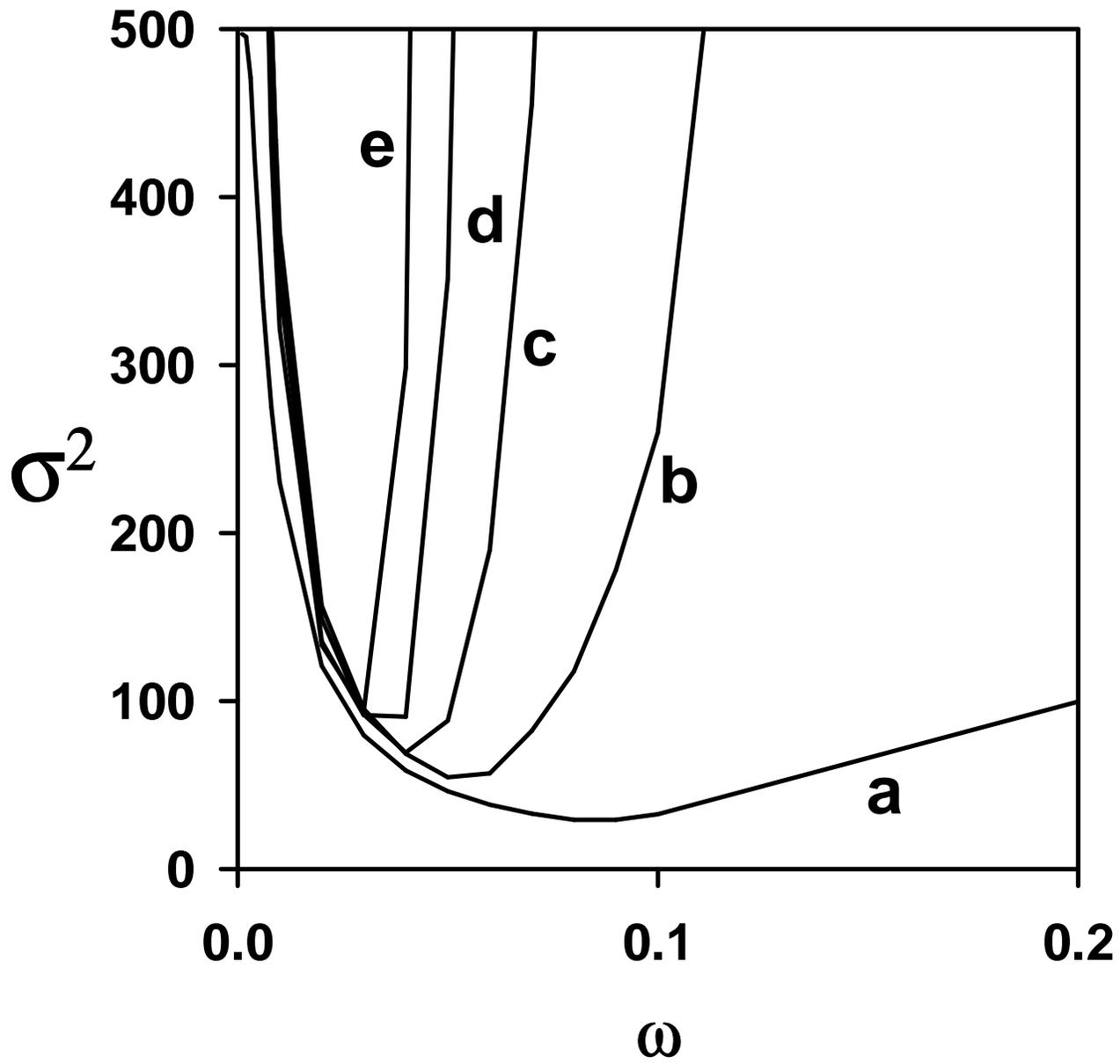
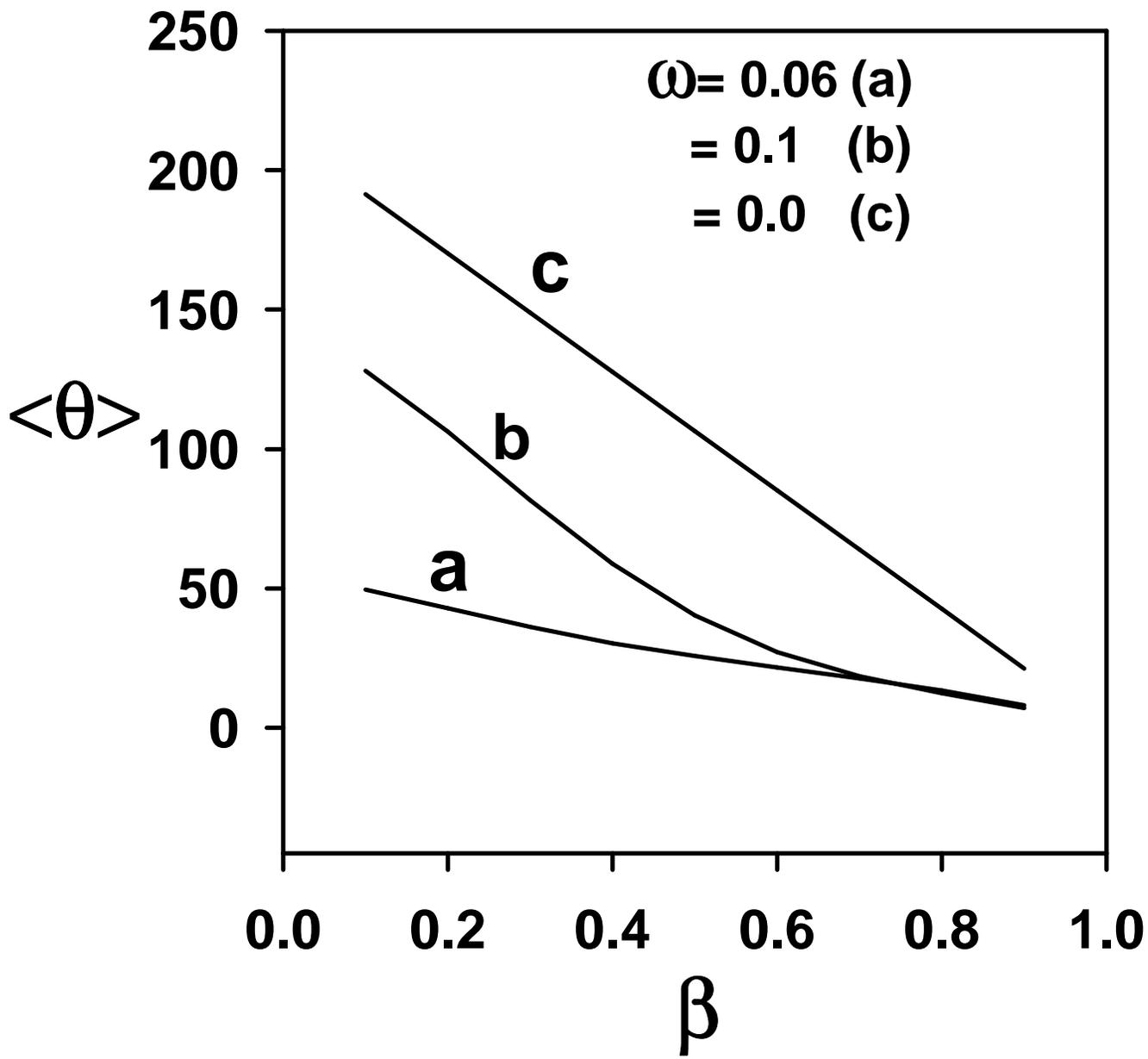
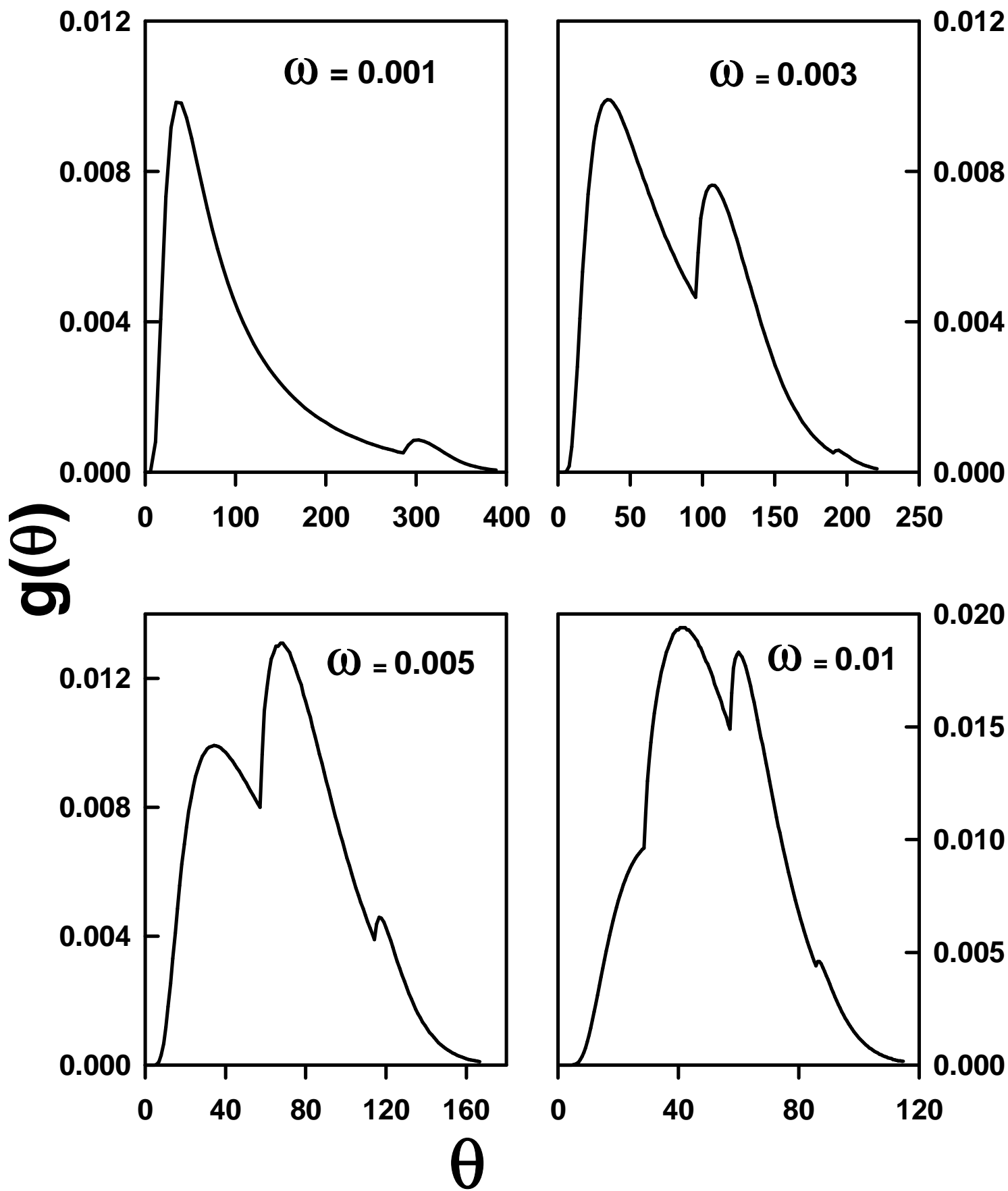
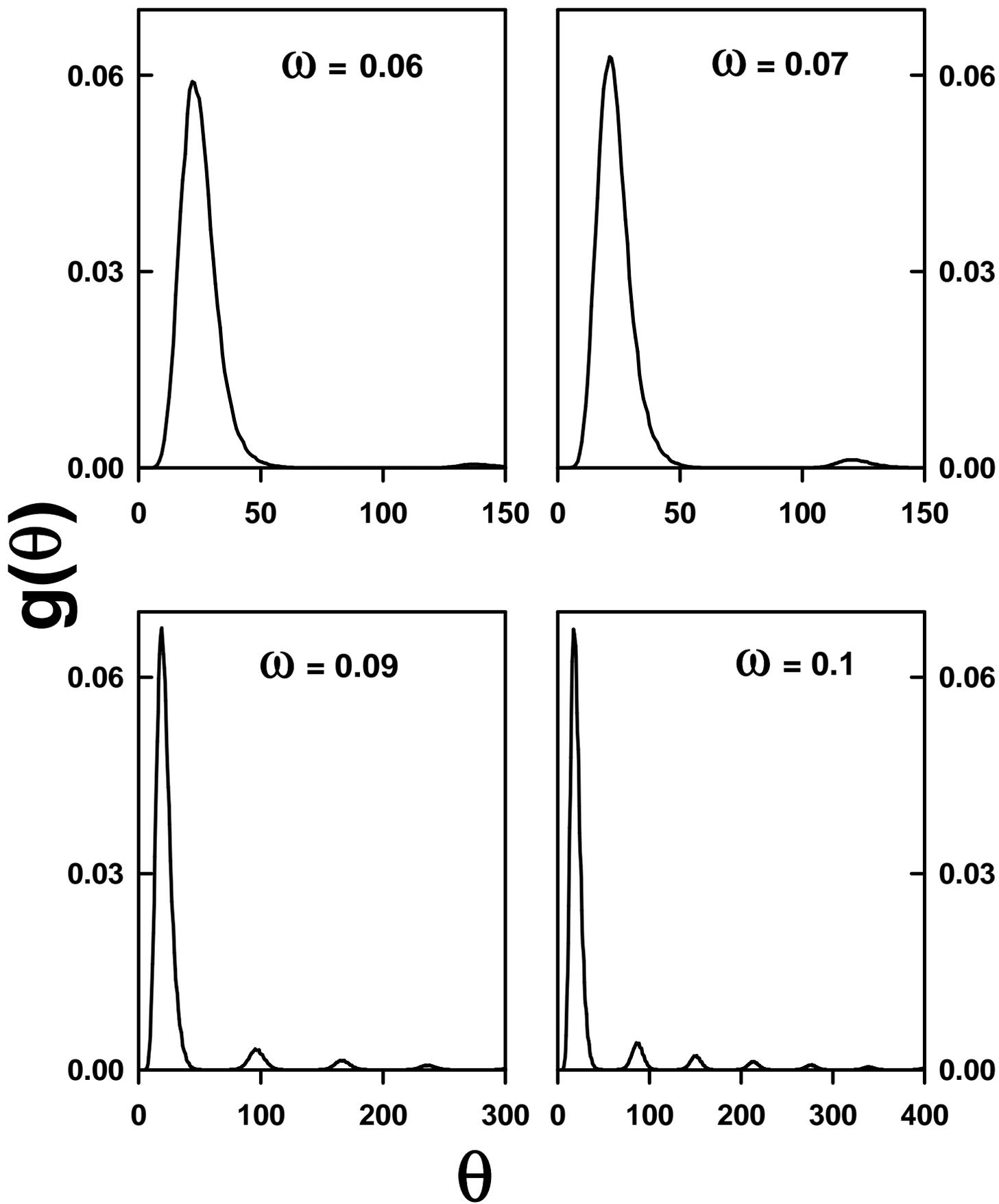
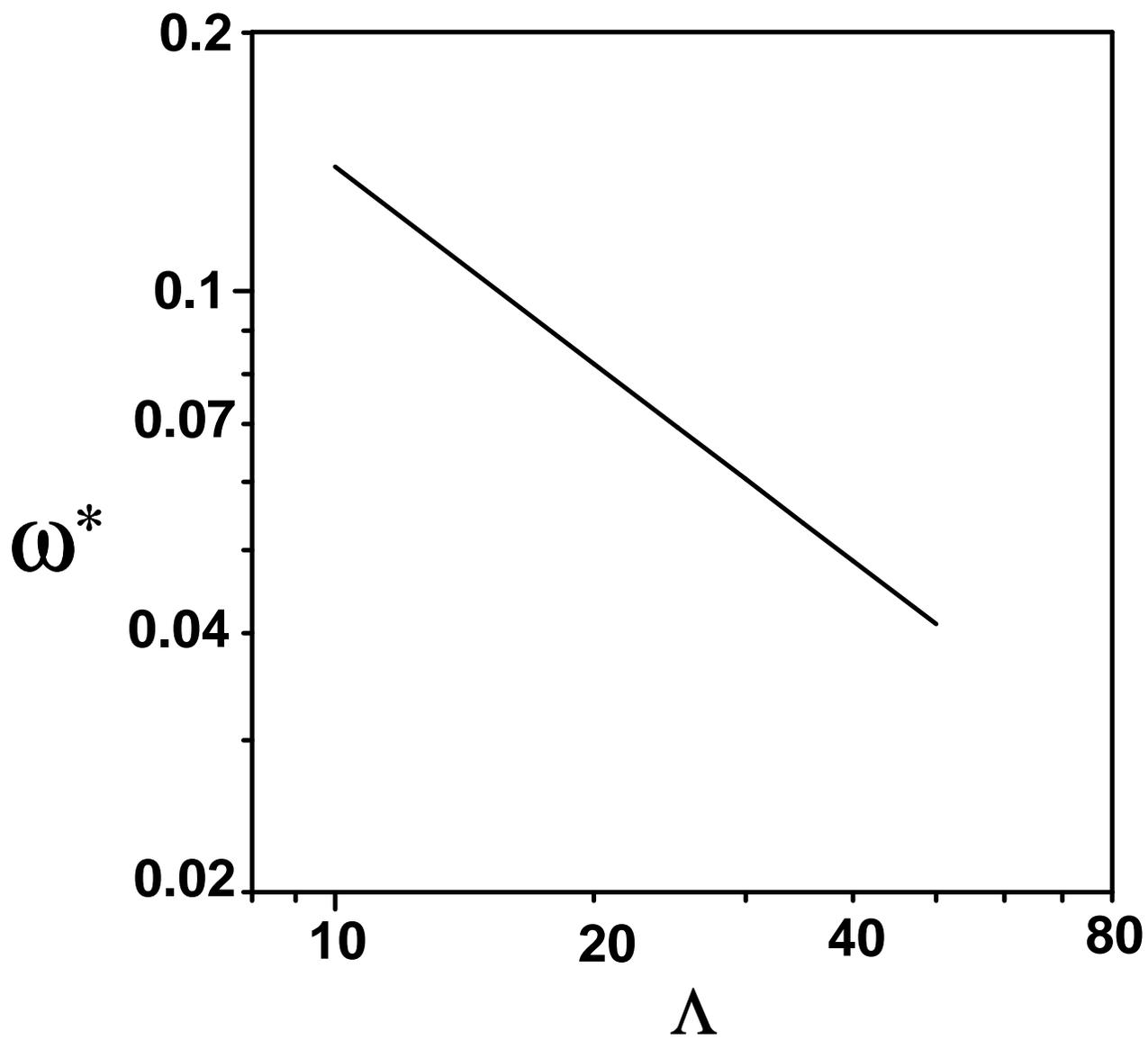
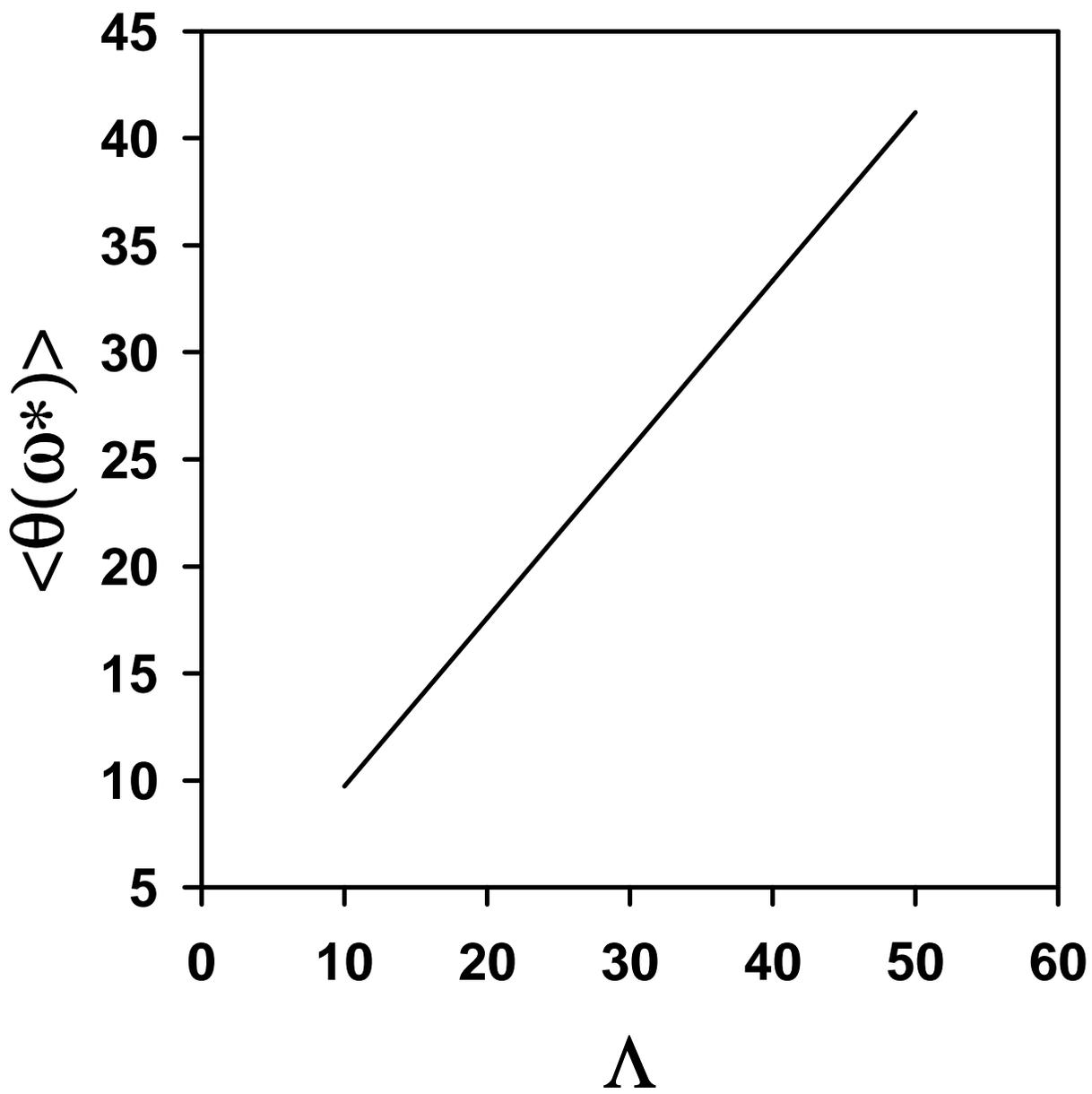
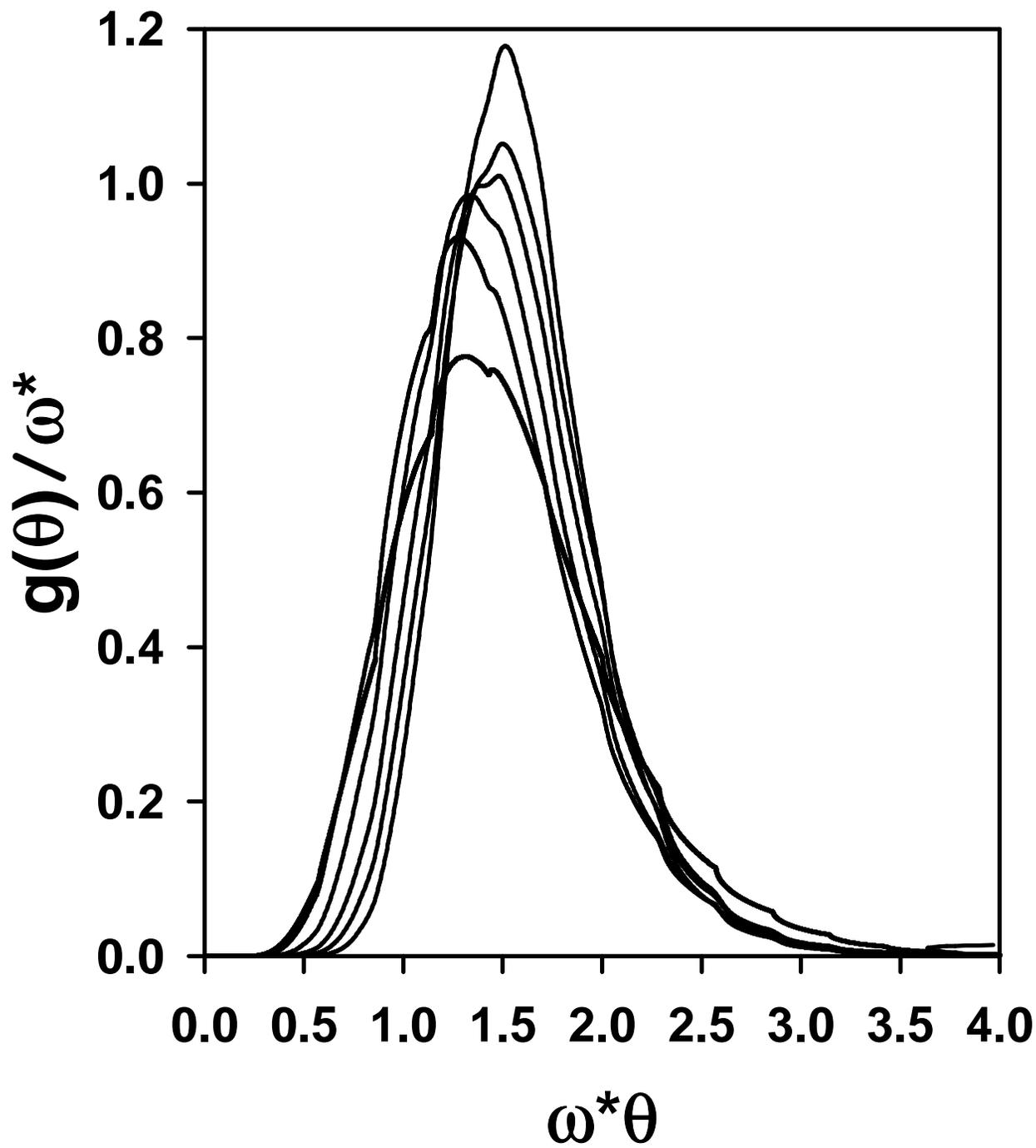
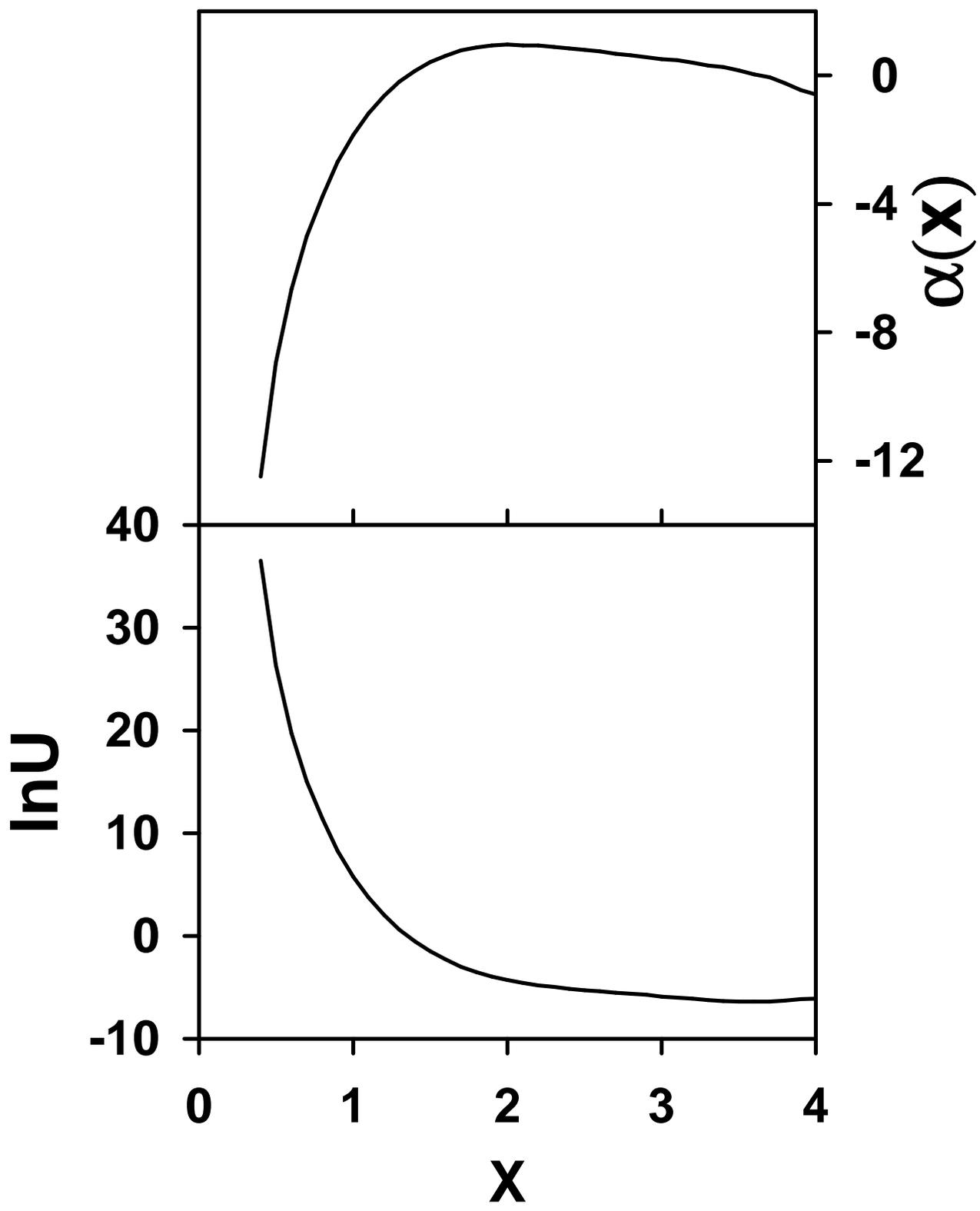
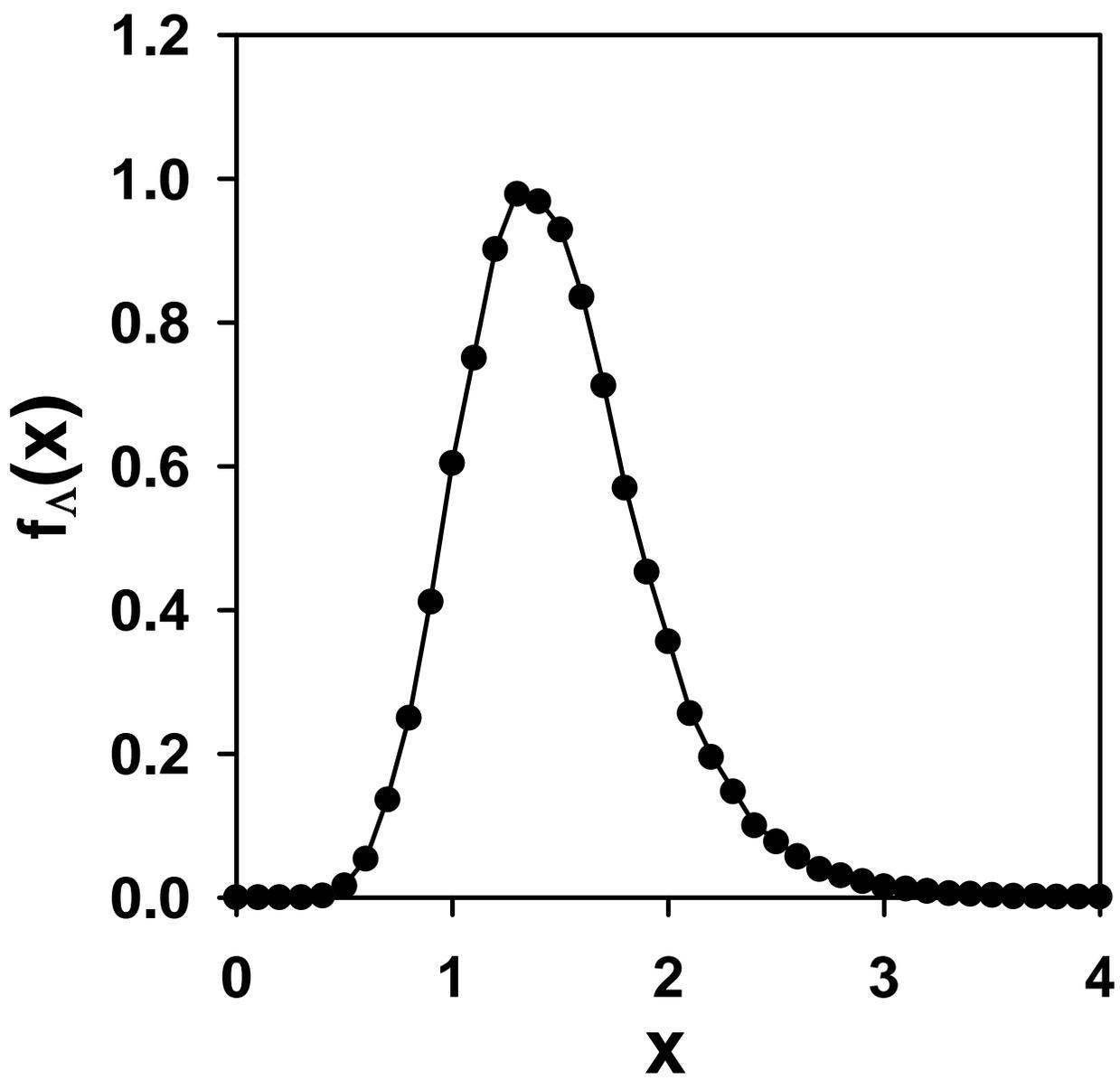
\end{document}